
\def\date{\ifcase\month\or
  January\or February\or March\or April\or
  May\or June\or July\or August\or
  September\or October\or November\or
  December\fi\space\number\day, \number\year}
%
\def\xxx{\vbox {\hbox{\lower 0.9\baselineskip \hbox{$<$}} \break
               \hbox{\lower 0.2\baselineskip \hbox{$\sim$}} } }
\def\newline{\hfill\break}
\def\boxed#1{\vrule\vbox{\hrule\kern2pt
    \hbox{\kern2pt #1\kern2pt}\kern2pt\hrule}\vrule}


\def\prb#1,#2,#3.{Phys. Rev. B {\bf #1} #2 (\kern-1.5pt#3).}
\def\prl#1,#2,#3.{Phys. Rev. Lett. {\bf #1} #2 (\kern-1.5pt#3).}


\let\endmode=\par               

\def\beginparmode{\endmode
  \begingroup \def\endmode{\par\endgroup}}

\def\body                      
  {\beginparmode}              
\def\endreferences{\par\endgroup}

\def\ref#1{Ref.[#1]}                   
\def\Ref#1{Ref.[#1]}                   

\def\references       
  {\immediate\write16{References}   
  {{\bf} REFERENCES}\par\bigskip          
   \beginparmode
   \frenchspacing \parindent=0pt \leftskip=1truecm
   \interlinepenalty=10000
   \parskip=8pt plus 3pt \everypar{\hangindent=\parindent}}
\def\refto#1{$^{(#1)}$}          
\gdef\refis#1{\indent\hbox to 0pt{\hss[#1]~}} 

\catcode`@=11
\newcount\r@fcount \r@fcount=0
\newcount\r@fcurr
\immediate\newwrite\reffile
\newif\ifr@ffile\r@ffilefalse
\def\w@rnwrite#1{\ifr@ffile\immediate\write\reffile{#1}\fi\message{#1}}

\def\writer@f#1>>{}
\def\referencefile{
  \r@ffiletrue\immediate\openout\reffile=\jobname.ref%
  \def\writer@f##1>>{\ifr@ffile\immediate\write\reffile%
    {\noexpand\refis{##1} = \csname r@fnum##1\endcsname = %
     \expandafter\expandafter\expandafter\strip@t\expandafter%
     \meaning\csname r@ftext\csname r@fnum##1\endcsname\endcsname}\fi}%
  \def\strip@t##1>>{}}

\def\citeall#1{\xdef#1##1{#1{\noexpand\cite{##1}}}}
\def\cite#1{\each@rg\citer@nge{#1}}	

\def\each@rg#1#2{{\let\thecsname=#1\expandafter\first@rg#2,\end,}}
\def\first@rg#1,{\thecsname{#1}\apply@rg}	
\def\apply@rg#1,{\ifx\end#1\let\next=\relax
\else,\thecsname{#1}\let\next=\apply@rg\fi\next}

\def\citer@nge#1{\citedor@nge#1-\end-}	
\def\citer@ngeat#1\end-{#1}
\def\citedor@nge#1-#2-{\ifx\end#2\r@featspace#1 
  \else\citel@@p{#1}{#2}\citer@ngeat\fi}	
\def\citel@@p#1#2{\ifnum#1>#2{\errmessage{Reference range #1-#2\space is bad.}%
    \errhelp{If you cite a series of references by the notation M-N, then M and
    N must be integers, and N must be greater than or equal to M.}}\else%
 {\count0=#1\count1=#2\advance\count1
by1\relax\expandafter\r@fcite\the\count0,%
  \loop\advance\count0 by1\relax
    \ifnum\count0<\count1,\expandafter\r@fcite\the\count0,%
  \repeat}\fi}

\def\r@featspace#1#2 {\r@fcite#1#2,}	
\def\r@fcite#1,{\ifuncit@d{#1}
    \newr@f{#1}%
    \expandafter\gdef\csname r@ftext\number\r@fcount\endcsname%
                     {\message{Reference #1 to be supplied.}%
                      \writer@f#1>>#1 to be supplied.\par}%
 \fi%
 \csname r@fnum#1\endcsname}
\def\ifuncit@d#1{\expandafter\ifx\csname r@fnum#1\endcsname\relax}%
\def\newr@f#1{\global\advance\r@fcount by1%
    \expandafter\xdef\csname r@fnum#1\endcsname{\number\r@fcount}}

\let\r@fis=\refis			
\def\refis#1#2#3\par{\ifuncit@d{#1}
   \newr@f{#1}%
   \w@rnwrite{Reference #1=\number\r@fcount\space is not cited up to now.}\fi%
  \expandafter\gdef\csname r@ftext\csname r@fnum#1\endcsname\endcsname%
  {\writer@f#1>>#2#3\par}}

\def\ignoreuncited{
   \def\refis##1##2##3\par{\ifuncit@d{##1}%
     \else\expandafter\gdef\csname r@ftext\csname
r@fnum##1\endcsname\endcsname%
     {\writer@f##1>>##2##3\par}\fi}}

\def\r@ferr{\endreferences\errmessage{I was expecting to see
\noexpand\endreferences before now;  I have inserted it here.}}
\let\r@ferences=\references
\def\references{\r@ferences\def\endmode{\r@ferr\par\endgroup}}

\let\endr@ferences=\endreferences
\def\endreferences{\r@fcurr=0
  {\loop\ifnum\r@fcurr<\r@fcount
    \advance\r@fcurr by 1\relax\expandafter\r@fis\expandafter{\number\r@fcurr}%
    \csname r@ftext\number\r@fcurr\endcsname%
  \repeat}\gdef\r@ferr{}\endr@ferences}


\let\r@fend=\endpaper\gdef\endpaper{\ifr@ffile
\immediate\write16{Cross References written on []\jobname.REF.}\fi\r@fend}

\catcode`@=12

\citeall\refto		
\citeall\ref		%
\citeall\Ref		%



  \font\twelverm=cmr10 scaled 1200       \font\twelvei=cmmi10 scaled 1200
  \font\twelvesy=cmsy10 scaled 1200      \font\twelveex=cmex10 scaled 1200
  \font\twelvebf=cmbx10 scaled 1200      \font\twelvesl=cmsl10 scaled 1200
  \font\twelvett=cmtt10 scaled 1200      \font\twelveit=cmti10 scaled 1200

  \font\twelvemib=cmmib10 scaled 1200
  \font\elevenmib=cmmib10 scaled 1095
  \font\tenmib=cmmib10
  \font\eightmib=cmmib10 scaled 800


    \font\eleveni=cmmi10 scaled 1095
\font\elevensy=cmsy10 scaled 1095


\font\seventeeni=cmmi10 scaled \magstep3

\font\seventeensy=cmsy10 scaled \magstep3

\font\seventeenmib=cmmib10 scaled \magstep3

\font\fourteenbf=cmbx10 scaled\magstep2


\newfam\cpfam%



\skewchar\eleveni='177   \skewchar\elevensy='60
\skewchar\elevenmib='177  \skewchar\seventeensy='60
\skewchar\seventeenmib='177
\skewchar\seventeeni='177

\newfam\mibfam%


  \skewchar\twelvei='177   \skewchar\twelvesy='60
  \skewchar\twelvemib='177
%
%
\def\twelvepoint{\normalbaselineskip=12.4pt
  \abovedisplayskip 12.4pt plus 3pt minus 9pt
  \belowdisplayskip 12.4pt plus 3pt minus 9pt
  \abovedisplayshortskip 0pt plus 3pt
  \belowdisplayshortskip 7.2pt plus 3pt minus 4pt
  \smallskipamount=3.6pt plus 1.2pt minus 1.2pt
  \medskipamount=7.2pt plus 2.4pt minus 2.4pt
  \bigskipamount=14.4pt plus 4.8pt minus 4.8pt
  \def\rm{\fam0\twelverm}          \def\it{\fam\itfam\twelveit}%
  \def\sl{\fam\slfam\twelvesl}     \def\bf{\fam\bffam\twelvebf}%
  \def\mit{\fam 1}                 \def\cal{\fam 2}%
  \def\tt{\twelvett}%
  \def\mib{\fam\mibfam\twelvemib}%

  \textfont0=\twelverm   \scriptfont0=\tenrm     \scriptscriptfont0=\sevenrm
  \textfont1=\twelvei    \scriptfont1=\teni      \scriptscriptfont1=\seveni
  \textfont2=\twelvesy   \scriptfont2=\tensy     \scriptscriptfont2=\sevensy
  \textfont3=\twelveex   \scriptfont3=\twelveex  \scriptscriptfont3=\twelveex
  \textfont\itfam=\twelveit
  \textfont\slfam=\twelvesl
  \textfont\bffam=\twelvebf
  \textfont\mibfam=\twelvemib       \scriptfont\mibfam=\tenmib
                                             \scriptscriptfont\mibfam=\eightmib

  \def\xrm{\textfont0=\twelverm\scriptfont0=\tenrm
      \scriptscriptfont0=\sevenrm\rm}
\normalbaselines\rm}

\catcode`@=11
\newcount\tagnumber\tagnumber=0

\immediate\newwrite\eqnfile
\newif\if@qnfile\@qnfilefalse
\def\write@qn#1{}
\def\writenew@qn#1{}
\def\w@rnwrite#1{\write@qn{#1}\message{#1}}
\def\@rrwrite#1{\write@qn{#1}\errmessage{#1}}

\def\taghead#1{\gdef\t@ghead{#1}\global\tagnumber=0}
\def\t@ghead{}

\expandafter\def\csname @qnnum-3\endcsname
  {{\t@ghead\advance\tagnumber by -3\relax\number\tagnumber}}
\expandafter\def\csname @qnnum-2\endcsname
  {{\t@ghead\advance\tagnumber by -2\relax\number\tagnumber}}
\expandafter\def\csname @qnnum-1\endcsname
  {{\t@ghead\advance\tagnumber by -1\relax\number\tagnumber}}
\expandafter\def\csname @qnnum0\endcsname
  {\t@ghead\number\tagnumber}
\expandafter\def\csname @qnnum+1\endcsname
  {{\t@ghead\advance\tagnumber by 1\relax\number\tagnumber}}
\expandafter\def\csname @qnnum+2\endcsname
  {{\t@ghead\advance\tagnumber by 2\relax\number\tagnumber}}
\expandafter\def\csname @qnnum+3\endcsname
  {{\t@ghead\advance\tagnumber by 3\relax\number\tagnumber}}

\def\equationfile{%
  \@qnfiletrue\immediate\openout\eqnfile=\jobname.eqn%
  \def\write@qn##1{\if@qnfile\immediate\write\eqnfile{##1}\fi}
  \def\writenew@qn##1{\if@qnfile\immediate\write\eqnfile
    {\noexpand\tag{##1} = (\t@ghead\number\tagnumber)}\fi}
}

\def\callall#1{\xdef#1##1{#1{\noexpand\call{##1}}}}
\def\call#1{\each@rg\callr@nge{#1}}

\def\each@rg#1#2{{\let\thecsname=#1\expandafter\first@rg#2,\end,}}
\def\first@rg#1,{\thecsname{#1}\apply@rg}
\def\apply@rg#1,{\ifx\end#1\let\next=\relax%
\else,\thecsname{#1}\let\next=\apply@rg\fi\next}

\def\callr@nge#1{\calldor@nge#1-\end-}
\def\callr@ngeat#1\end-{#1}
\def\calldor@nge#1-#2-{\ifx\end#2\@qneatspace#1 %
  \else\calll@@p{#1}{#2}\callr@ngeat\fi}
\def\calll@@p#1#2{\ifnum#1>#2{\@rrwrite{Equation range #1-#2\space is bad.}
\errhelp{If you call a series of equations by the notation M-N, then M and
N must be integers, and N must be greater than or equal to M.}}\else%
 {\count0=#1\count1=#2\advance\count1
 by1\relax\expandafter\@qncall\the\count0,%
  \loop\advance\count0 by1\relax%
    \ifnum\count0<\count1,\expandafter\@qncall\the\count0,%
  \repeat}\fi}

\def\@qneatspace#1#2 {\@qncall#1#2,}
\def\@qncall#1,{\ifunc@lled{#1}{\def\next{#1}\ifx\next\empty\else
  \w@rnwrite{Equation number \noexpand\(>>#1<<) has not been defined yet.}
  >>#1<<\fi}\else\csname @qnnum#1\endcsname\fi}

\let\eqnono=\eqno
\def\eqno(#1){\tag#1}
\def\tag#1$${\eqnono(\displayt@g#1 )$$}

\def\aligntag#1\endaligntag
  $${\gdef\tag##1\\{&(##1 )\cr}\eqalignno{#1\\}$$
  \gdef\tag##1$${\eqnono(\displayt@g##1 )$$}}

\def\eqalignno#1{\displ@y \tabskip\centering
  \halign to\displaywidth{\hfil$\displaystyle{##}$\tabskip\z@skip
    &$\displaystyle{{}##}$\hfil\tabskip\centering
    &\llap{$\displayt@gpar##$}\tabskip\z@skip\crcr
    #1\crcr}}

\def\displayt@gpar(#1){(\displayt@g#1 )}

\def\displayt@g#1 {\rm\ifunc@lled{#1}\global\advance\tagnumber by1
        {\def\next{#1}\ifx\next\empty\else\expandafter
        \xdef\csname @qnnum#1\endcsname{\t@ghead\number\tagnumber}\fi}%
  \writenew@qn{#1}\t@ghead\number\tagnumber\else
        {\edef\next{\t@ghead\number\tagnumber}%
        \expandafter\ifx\csname @qnnum#1\endcsname\next\else
        \w@rnwrite{Equation \noexpand\tag{#1} is a duplicate number.}\fi}%
  \csname @qnnum#1\endcsname\fi}

\def\ifunc@lled#1{\expandafter\ifx\csname @qnnum#1\endcsname\relax}

\let\@qnend=\end\gdef\end{\if@qnfile
\immediate\write16{Equation numbers written on []\jobname.EQN.}\fi\@qnend}

\catcode`@=12
\def\Eq#1{Eq. (\call{#1})}                 

\twelvepoint
\tolerance=2000
\baselineskip=0.8truecm
\centerline{\fourteenbf Fermi-surface properties of the }
\centerline{\fourteenbf perovskite superconductors}
\bigskip\bigskip
\centerline{\bf Claudius Gros, Roser Valent\'\i}
\bigskip
\centerline{\it Institut f\"ur Physik, Universit\"at Dortmund,}
\centerline{\it 44221 Dortmund, Germany,}
\centerline{\it email: UPH301 and UPH084 at DDOHRZ11}
\bigskip\bigskip\bigskip

We calculate Fermi-surface properties of the Cuprate
superconductors within the three-band Hubbard model using a
cluster expansion for the proper self-energy. The
Fermi-surface topology is in agreement with
angular-resolved photoemission data for dopings
$\sim 20\%$. We discuss possible violations
of the Luttinger sum-rule for smaller dopings
and the role of van-Hove singularities in the
density of  states of the
Zhang-Rice singlets. We calculate the shift
in the chemical potential
upon doping and find quantitative
agreement with recent experiments.

\bigskip




\vfill\eject

%
%
%

\baselineskip0.55truecm

\bigskip
{\bf Introduction}
In the last years, the experimental characterization
of Fermi-surface properties of the high-temperature
superconductors has progressed with the help
of angle-resolved\refto{effective_mass_2,effective_mass_4,
Fermi_surface,Liu_et_al,Tobin_et_al}
and angle-integrated\refto{Veenendaal_S_S_G}
photoemission studies. The such determined
Fermi-surface topology agrees quite well with
results from band-structure calculations based
on the local-density approximation\refto{band_structure}.
This measured Fermi-surface topology has been taken
as an input for theoretical
calculations\refto{K_Levin} in the so called
``Fermiology'' scenario.
Many theorists use, on the other hand, a simplified
three-band Hamiltonian (also called
Emery model\refto{Emery}), which incorporates
strong, local correlations, as exemplified by the
composite-operator approach\refto{composite_operator}.
Determination of the Fermi-surface properties of the
Emery model is of central interest and has been investigated
in a series of recent studies
\refto{slave_boson,Fulde,hole_pockets,Poilblanc_Z_S_D}.

It is difficult to calculate with
numerical methods, like exact
diagonalization\refto{Eskes_Sawatzky,Ohta_T_K_S_M}
or Quantum Monte-Carlo\refto{Dopf_W_D_M_H},
the Fermi-surface topology for the three-band model
directly, due to cluster-size limitations.
The numerical methods have been shown, on the other side,
to be very useful for extracting
the energy scales of the local
charge- and spin excitations\refto{Eskes_Sawatzky}.
Here we propose that the {\it cluster self-energy}
provides the missing link between such numerical
approaches and calculation of momentum-dependent
properties, like the Fermi-surface topology.

%
%
%

\bigskip
{\bf Method}
The self-energy of finite
clusters with open boundary conditions
contains in a very precise way
information upon the Green's function and self-energy
of the {\it extended} system\refto{Gros_Valenti_93}.
Let us consider the
real-space components of the self-energy for both the
extended system, $\Sigma_{i,j}(\omega)$, and for a given
finite cluster with open boundary conditions,
$\Sigma_{i,j}^{(c)}(\omega)$. It is
straightforward,
that all irreducible diagrams contributing to
$\ \Sigma_{i,j}^{(c)}(\omega)\ $ are a subset of the
diagrams contributing to $\ \Sigma_{i,j}(\omega)$.
Therefore the cluster self-energy is a {\it systematic}
approximation to the self-energy of the full system.

Typically one first evaluates the cluster
Green's function, $G_{i,j}^{(c)}(\omega)$, in an exact
diagonalization study. The cluster self-energy is
then obtained by inverting the cluster Dyson's equation
$$
G_{i,j}^{(c)}(\omega) \ =\
G_{i,j}^{(c,0)}(\omega) +
\sum_{m,n} G_{i,n}^{(c,0)}(\omega)
       \Sigma_{m,n}^{(c)}(\omega)
            G_{n,j}^{(c)}(\omega),
\eqno(cluster_Dyson)
$$
where the $\ i,m,n,j\ $ are sites of the cluster and
$\ G_{i,j}^{(c,0)}(\omega)\ $ is the non-interacting
cluster Green's function.

The Green's function of the full CuO$_2$ lattice
is determined by inserting the $\ \Sigma_{m,n}^{(c)}(\omega)\ $
determined from \Eq{cluster_Dyson}, into the
Dyson's equation of the extended system
$$
G_{i,j}^{}(\omega) \ =\
G_{i,j}^{(0)}(\omega) +
\sum_{m,n} G_{i,n}^{(0)}(\omega)
       \Sigma_{m,n}^{(c)}(\omega)
            G_{n,j}^{}(\omega),
\eqno(full_Dyson)
$$
where the $\ i,m,n,j\ $ run over all sites of the CuO$_2$
lattice and $\ G_{i,j}^{(0)}(\omega)\ $ is the non-interacting
Green's function of the CuO$_2$ plane. The Green's function,
$G_{i,j}(\omega)$, obtained by \Eq{full_Dyson} can be
systematically improved considering larger and larger
clusters in \Eq{cluster_Dyson}. The smallest cluster one
could possibly consider is the one-site-cluster and
\Eq{full_Dyson} together with \Eq{cluster_Dyson} would then
reduce to the Hubbard-I approximation\refto{Hubbard}.
For the Emery model we consider the CuO$_4$ cluster
for \Eq{cluster_Dyson}. Within this approximation both the
important local charge fluctuations and the singlet-triplet
splitting are included.

The passage of the
quasi-particle peak through the chemical potential,
determined from the {\bf k}-dependent
Green's function, $\ G_{\bf k}(\omega)\equiv 1/L\sum_{i,j}
\exp[i{\bf k}({\bf R}_i-{\bf R}_j)] G_{i,j}(\omega)\ $
determines then the Fermi-surface. The density of electrons is
given by
$$
n\equiv 2\int_{-\infty}^{\infty}d\omega
      ({-1\over\pi}) Im G(\omega)
    n_F(\omega-\mu)
 = 2\int_{{\bf k}\in FS}
  {d^2k\over(2\pi)^2} Z_{\bf k},
\eqno(density)
$$
with $n_F(\omega-\mu)$ and $Z_{\bf k}$ being the
Fermi distribution function and the
quasiparticle spectral weight respectively.
The factor of two in \Eq{density} comes from
spin-degeneracy and the {\bf k}-sum is over all occupied
states. The density of electrons, as given by \Eq{density},
is to be
determined self-consistently\refto{Gros_Valenti_93}
as a function of the chemical potential, $\mu$,
by demanding it to be equal to the average
density of electrons of the CuO$_4$ cluster, for which
$\ \Sigma_{i,j}^{(c)}(\omega)\ $ is determined via
\Eq{cluster_Dyson}.

%
%
%

\bigskip
{\bf Results}
For the three-band Hamiltonian we have
chosen (in electron notation), following
Eskes and Sawatzky\refto{Eskes_Sawatzky},
$\epsilon_p-\epsilon_d=5.3$eV for the
difference in oxygen- and copper orbital
energies\refto{hole_notation}, $t_{dp}=1.3$eV and $t_{pp}=0.65$eV
for the Cu-O and the O-O hopping matrix element
respectively and $U_d=8.8$eV for the onsite Coulomb
repulsion on the copper site.

{}From the Green's function of a CuO$_4$ cluster we have
extracted the self-energy and used it as an approximation
for the self-energy of the
infinite CuO$_2$ layer, as explained above. In
Fig.1 we illustrate typical results for the angle-integrated
spectral function and a doping $x\sim20\%$. For comparison
in Fig.1(c) the density of states of the non-interacting
CuO$_2$ plane (i.e. with $U_d=0$) is shown.
The bonding-band (mainly copper) is located around $\sim-9$eV, the
antibonding band (mainly oxygen) extends from $\sim-3$eV
to $\sim1$eV and the non-bonding oxygen orbitals is in between.
For every band the typical 2D logarithmic van-Hove singularity
is clearly visible, for the non-bonding band,
for instance, it is located
at $\mu-5.3$eV$\ =-3.1$eV. In Fig.1(b) the  location and the
weight of the peaks of the CuO$_4$-cluster Green's function
are shown.

The results for the full Green's function of the
extended CuO$_2$ layer are presented in Fig.1(a).
Coherently propagating many-body bands have formed out of
the cluster energy-levels\refto{Gros_Valenti_93}.
The chemical potential lies in the
band of Zhang-Rice singlets\refto{Zhang_Rice},
separated by a charge-transfer gap
of about $\sim 1$eV from the empty band of copper-d$^{10}$ states.
The non-bonding oxygen orbitals and the band of Zhang-Rice
triplets are located  -3eV to -5.5eV below the Fermi-level.
\smallskip
%
%
%

{\bf Van Hove singularities}
A van Hove singularity can be observed in the band of
Zhang-Rice singlets (see Fig.1(a)). It is tempting
to identify this structure with the very sharp feature
observed in recent angle-resolved photoemission
experiments\refto{Tobin_et_al}
on untwinned YBa$_2$Cu$_3$O$_{6.9}$ about 1eV below the
Fermi-surface. Let us note that, contrary to predictions
based on a weak-coupling picture, the van Hove singularity
in the density of states does not occur at half-filling in
  our many body calculation.
  We find that only for relative large
dopings, namely of about $38\%$, the
Fermi-surface would be located right at the
van Hove singularity. The absence of a van Hove
singularity at the Fermi-edge for dopings $x<0.2$
is consistent with recent NMR experiments\refto{Song_K_P_H}
and in  contrast to a band-structure calculation
based on the local density approximation\refto{Xu_et_al}.
 Also we note that these results do not support the recently proposed
``van Hove scenario'' of cuprate
supercoductivity\refto{van_Hove_super}.
%
%
%

\smallskip
{\bf Fermi surface}
In Fig.2(a) we present the results for the Fermi-surface
as a function of doping, $x=0.01,\dots,0.70$. Also given
are values of the quasiparticle spectral weight,
$Z_{\bf k}$, along the (1,1) direction at the Fermi edge.
In Fig.2(b) we present, for comparison, the Fermi-surface of the
non-interacting CuO$_2$ layer at various dopings. (0,0)
corresponds to the $\Gamma$-point.

We find that the Fermi-surface presented in
Fig.2(a) is in agreement with angular resolved
photoemission
experiments\refto{effective_mass_2,Fermi_surface}
for dopings of about $x\sim20\%$.
The location along the (1,0)---(1,1) lines does also
agree quantitatively quite well. Along the (0,0)---(1,1)
line the calculated Fermi edge is located at somewhat too
large momenta.
We note that the quasiparticle spectral weight is
reduced by a factor 2 to 5, depending on the doping.
For small dopings the calculated Fermi surface expands
rapidly, due to the reduced spectral weight, $Z_{\bf k}$
(compare \Eq{density}) and finally closes completely
for $x=0$ at the (1,1) point. Similar results for the
volume of the Fermi-sea as a function of doping
has been found by Unger and
Fulde\refto{Fulde} with the projection technique.
Note that the data presented
in Fig.1 are for a paramagnetic state and are therefore
not in contradiction
with the notion\refto{hole_pockets,Stephan_Horsch_91}
that in the antiferromagnetic state and dopings of a
few percent, the Fermi-surface
should take the form of ``hole-pockets'' centered around
$(\pi/2,\pi/2)$.

For small interaction strength, in perturbation theory,
the Luttinger sum-rule\refto{Luttinger} is valid.
It states, that the volume of the Fermi-sea of the
system with interaction should be equal to that of
the non-interacting system.
Comparing Fig.2(b) and Fig.2(a) we note that
the volume of the Fermi sea obtained for $\ U_d=8.8$eV
is larger than the one predicted by the Luttinger sum-rule
(i.e. those obtained for $U_d=0.0$). E.g. for
$\ x=20\%\ $ the difference is $\sim14\%$ of the
Brillouin zone.
The cluster-expansion for the self-energy
used here is best for large interacting strength and
it is therefore not so surprising that the results for
the Fermi-surface violate the Luttinger sum-rule, which
may be expected to hold for small interactions,
$U_d$. Let us note, that for very large dopings
the Fermi-surface for both $\ U_d=8.8$eV and $\ U_d=0$ have nearly
identical volumes (for $x=70\%$ they agree within a
percent), since the doped charge carriers
go predominantly into the oxygen orbitals, which
are influenced only indirectly by $\ U_d$.

Quantitatively the Luttinger sum-rule
can be experimentally verified
best by comparison with results from
band-structure calculations.
The experimentally determined Fermi-surfaces
(accuracy\refto{Liu_et_al}: 10-20\% of the
Brillouin zone) agree quite well
with band-structure calculations\refto{band_structure}.
Since, on the other hand, standard
band-structure calculations employ only one
Slater determinant they necessarily fulfill the
Luttinger sum-rule (and yield a $Z_{\bf k}\equiv1$).
%
%
%

\smallskip
{\bf Chemical potential}
In Fig.3 we present the shift
(relative to half-filling\refto{chem_pot})
in the chemical potential, $\mu$, upon doping, $x$.
For comparison the shift
in the chemical potential for the non-interacting
CuO$_2$ plane, $\mu_0$, is also shown, they
differ by about a factor of two. We have
included in Fig.3 the data obtained by a recent
photoemission study\refto{Veenendaal_S_S_G} on
Bi$_2$Sr$_2$Ca$_{1-x}$Y$_x$Cu$_2$O$_{8+\delta}$.
  In order to compare with the experimental results we
   adjusted our calculated position of the chemical
potential for $x=0$ to the experimental value for $x=0.01$.
Note that, eventhough the experimental data for $x=0.01\dots0.03$
 correspond to an insulator state, while we have a metallic state,
 excluding these points and adjusting at values, e.g., $x=0.08$,
 wouldn't change the curves significantly.

We also show in Fig.3 the doping dependence of the
renormalized Fermi-velocity, $v_F/v_{F0}\sim m/m^*$.
We find, for about $x=20\%$ doping, an effective
mass $m^*/m$ of about two, in agreement with
results from angle-resolved
photoemission experiments\refto{effective_mass_2}
and quantum Monte-Carlo studies\refto{Dopf_W_D_M_H}.
The effective mass is found to diverge for
dopings $\ x\rightarrow0$ (see Fig.3), as a signal of
the metal-insulator transition at half filling.
%
%
%

\smallskip
{\bf Conclusions}
It is still controversial how the Fermi surface of 2D
correlated, translational invariant electrons on a lattice
changes when going from the metal to the Mott-Hubbard insulator
regime as a function of electron density. While it seems
clear\refto{hole_pockets}
that the first holes doped into the half-filled,
antiferromagnetic ordered insulator, occupy states with
momenta around $(\pm\pi/2,\pm\pi/2)$, the situation for
finite doping concentrations in the paramagnetic state
is not so clear. Here we proposed a scenario, where the
volume of the Fermi-sea expands continuously with particle
density towards the $(\pi,\pi)$ point at half-filling.

For moderate dopings, $x\sim0.2\%$, this scenario is compatible
with exact diagonalization results of
one-band models\refto{Stephan_Horsch_91}
and angle-resolved photoemission\refto{Fermi_surface}
experiments on the cuprate superconductors. For
small dopings, $x=5-10\%$ we predict deviations from
the Luttinger sum-rule, which might be measurable.
The calculated shift in chemical potential associated with
the Fermi-surface topology is in reasonable agreement
with recent photoemission experiments\refto{Veenendaal_S_S_G}.

\bigskip
\line{\hfill * * * \hfill}
\bigskip

We would like to thank A. Kampf, R. Manzke,
G.A. Sawatzky, W. Weber and
W. Wenzel for discussions.
This work was supported by the Deutsche Forschungsgemeinschaft and
by the Minister f\"ur Wissenschaft
und Forschung des Landes Nordrhein-Westfalen.

\vfill\eject

%
%
%
\references

\refis{band_structure} M.S. Hybertsen, L.F. Mattheiss,
                       Phys. Rev. Lett. {\bf 60}, 1161 (1988);
                       W.E. Pickett,
                       Rev. Mod. Phys. {\bf 61}, 433 (1989);
                       W.E. Pickett, R.E. Cohen, H. Krakauer,
                       Phys. Rev. B {\bf 42}, 8764 (1990).

\refis{composite_operator} H. Matsumoto,  M. Sasaki,
                           S. Ishihara, M. Tachiki,
                           Phys. Rev. B {\bf 46}, 3009 (1992);
                           M. Sasaki, H. Matsumoto, M. Tachiki,
                           {\it ibid.}, 3022 (1992).

\refis{chem_pot} The limit
                 $\lim_{T\rightarrow0}\lim_{x\rightarrow0}
                  \mu(x,T)$ for the
                 doping and temperature dependent
                 chemical potential is the middle of the
                 charge transfer-gap. Within our zero-temperature
                 formalism the position of the chemical potential
                 at half-filling,
                 $\lim_{x\rightarrow0}\lim_{T\rightarrow0}
                  \mu(x,T)$, coincides with the
                 top of the valence band.

\refis{Dopf_W_D_M_H} G. Dopf, J. Wagner, P. Dietrich,
                     A. Muramatsu, W. Hanke,
                     Phys. Rev. Lett. {\bf 68}, 2082 (1992).

\refis{effective_mass_2} C.G. Olson {\it et al.},
                         Phys. Rev. B {\bf 42}, 381 (1990).

\refis{effective_mass_4} G. Mante {\it et al.},
                         Zeit. Phys.  {\bf 80}, 181 (1990).

\refis{Emery} V.J. Emery, Phys. Rev. Lett. {\bf 58}, 2794 (1987).

\refis{Eskes_Sawatzky} H. Eskes, L.H. Tjeng, G.A. Sawatzky,
                       Phys. Rev. B {\bf 41}, 288 (1990);
                       H. Eskes, G.A. Sawatzky,
                       Phys. Rev. B {\bf 44}, 9556 (1991).

\refis{Fermi_surface} J.C. Campuzano {\it et al.},
                      Phys. Rev. Lett. {\bf 64}, 2308 (1990).

\refis{Fulde} P. Unger, P. Fulde,
              {\it Spectral Function of Holes in the Emery
              Model}, to be published.


\refis{Gros_Valenti_93} C. Gros, R. Valent\'\i,
                        Phys. Rev. B {\bf 48}, 418 (1993).

\refis{hole_notation} In hole-notation the orbital energy
                      difference would be
                      $U_d-5.3$eV$\ =\ 3.5$eV.

\refis{Hubbard} J. Hubbard,
                Proc. R. Soc. A {\bf 276}, 238 (1963);
                {\bf 277}, 238 (1964);
                {\bf 281}, 401 (1964).

\refis{hole_pockets} C.L. Kane,  P.A. Lee, N. Read,
                     Phys. Rev. B {\bf 39}, 6880 (1989);
                     M.D. Johnson, C. Gros, K.J. von Szczepanski,
                     Phys. Rev. B {\bf 43}, 11 207 (1991).


\refis{Liu_et_al} Liu {\it et al.},
                  Phys. Rev. B {\bf 46}, 11 056 (1992).

\refis{Luttinger} J. Luttinger, J. Ward,
                  Phys. Rev. {\bf 118}, 1417 (1960);
                  J. Luttinger,
                  Phys. Rev. {\bf 119}, 1153 (1961);
                  G. Benefatto, G. Gallavotti,
                  J. Stat. Phys. {\bf 59}, 541 (1990).

\refis{K_Levin} See Q. Si, J.P. Lu, K. Levin
                Phys. Rev. B {\bf 45}, 4930 (1992)
                and references therein.

\refis{Ohta_T_K_S_M} Y. Ohta, K. Tsutsui, W. Koshibae,
                     T. Shimozato, S. Maekawa,
                     Phys. Rev. B {\bf 46}, 14 022 (1992).

\refis{Poilblanc_Z_S_D} D. Poilblanc, T. Ziman, H.J. Schulz,
                        E. Dagotto,
                        Phys. Rev. B {\bf 47}, 14 267 (1993).

\refis{slave_boson} J. Schmalian, G. Baumg\"artel, K.-H. Bennemann,
                    Phys. Rev. Lett. {\bf 68}, 1406 (1992).

\refis{Song_K_P_H} Y.-Q. Song, M.A. Kennard, K.R. Poppelmeier,
                   W.P. Halperin,
                   Phys. Rev. Lett. {\bf 70}, 3131 (1993).


\refis{Stephan_Horsch_91} W. Stephan, P. Horsch
                          Phys. Rev. Lett. {\bf 66}, 2258 (1991).

\refis{Tobin_et_al} J.G. Tobin {\it et al.},
                    Phys. Rev. B {\bf 45}, 5563 (1993).

\refis{van_Hove_super} D.M. News, C.C. Tsuei, P.C. Pattnaik,
                       C.L. Kane,
                       Comm. Cond. Mat. Phys. {\bf 15}, 273 (1992);
                       D.M. News, H.R. Krishnamurthy,
                       P.C. Pattnaik, C.C. Tsuei, C.L. Kane,
                       Phys. Rev. Lett. {\bf 69}, 1264 (1992).

\refis{Veenendaal_S_S_G} M.A. van Veenendaal, R. Schlatmann,
                         G.A. Sawatzky, W.A. Groen,
                         Phys. Rev. B {\bf 47}, 446 (1993).

\refis{Xu_et_al} J.-H. Xu, T.J. Waston-Yang, J. Yu, A.J. Freeman,
                 Phys. Lett. A {\bf 120}, 489 (1987) predict a
                 van Hove singularity at the Fermi surface for a
                 doping $x=0.17$ for La$_{2-x}$Sr$_x$CuO$_4$.

\refis{Zhang_Rice} F.C. Zhang, T.M. Rice,
                   Phys. Rev. B {\bf 37}, 3759 (1988).

\endreferences

\vfill\eject

{\bf Figure captions}
\bigskip\bigskip

\item{Fig.1} (a) A typical result for the angle-integrated
             spectal weight, for a doping x=0.2. The
             Fermi surface lies in the band of coherently
             propagating Zhang-Rice singlets.
             (b) The position and the weight of the poles
             of the CuO$_4$ cluster Green's function used
             for calculating the self-energy.
             (c) For comparison the density of states of the
             non-interacting ($U_d=0$) CuO$_2$ plane with
             the three van Hove singularities for the
             bonding, non-bonding and the anti-bonding band.

\bigskip

\item{Fig.2} (a) Results for the position of the Fermi surface
             in the first Brillouin zone, as a function of
             doping, $x$, and the quasiparticle renormalization
             factor, $Z$, at the Fermi edge. Note the strong
             reduction of $Z$ near the $(\pi,\pi)$ point.
             (b) For comparison the Fermi surface of the
             non-interacting ($U_d=0$) CuO$_2$ plane, which
             fulfills the Luttinger sum-rule, is shown.

\bigskip

\item{Fig.3} {\it Filled circles}: Results for the
             shift in the chemical potential, $\mu$,
             upon doping, $x$, relative to half-filling.
             {\it Crosses}: Experimental\refto{Veenendaal_S_S_G}
             results for the shift in chemical potential for
             Bi$_2$Sr$_2$Ca$_{1-x}$Y$_x$Cu$_2$O$_{8+\delta}$.
             {\it Open circles}: For comparison the shift in
             $\mu_0$ of the non-interacting, $U_d=0$, CuO$_2$
             plane.
             {\it Filled squares}: The renormalized Fermi
             velocity, $v_F/v_{F0}\sim m/m^*$. Note the
             reduction of about two for $x=0.2$.

\bye